\documentclass[twocolumn,amsmath,amssymb,prl]{revtex4}
\usepackage{graphicx}
\usepackage{dcolumn}
\usepackage{bm}

\begin{document}
\input{epsf}

\title[Observational Test for the Anthropic Argument] {An Observational
Test for the Anthropic Origin of the Cosmological Constant}

\author{Abraham Loeb}
\affiliation{Astronomy Department, Harvard University, 60 Garden Street, 
Cambridge, MA 02138, USA}

\begin{abstract}
The existence of multiple regions of space beyond the observable Universe
(within the so-called {\it multiverse}) where the vacuum energy density
takes different values, has been postulated as an explanation for the low
non-zero value observed for it in our Universe. It is often argued that our
existence pre-selects regions where the cosmological constant is
sufficiently small to allow galaxies like the Milky Way to form and
intelligent life to emerge.  At first glance, it would seem necessary to
visit regions of space far beyond our current horizon in order to
critically examine the validity of this {\it anthropic} argument.  However,
here we propose a simple empirical test for it within the boundaries of the
observable Universe.  We make use of the fact that dwarf galaxies formed in
our Universe at redshifts as high as $z\sim 10$ when the mean matter
density was larger by a factor of $\sim 10^3$ than today.  Existing
technology enables to check whether planets form in nearby dwarf galaxies
and globular clusters by searching for microlensing or transit events of
background stars.  The oldest of these nearby systems may have formed at
$z\sim 10$. Direct observations of dwarf galaxies at redshifts $z\sim 10$
can be used to characterize their size, mass, metallicity, and star
formation history, and identify the nearby systems that descended from
them.  If planets are as common per stellar mass in these descendents as
they are in the Milky Way galaxy, then the anthropic argument would be
weakened considerably since planets could have formed in our Universe even
if the cosmological constant, $\rho_V$, was three orders of magnitude
larger than observed.  For a flat probability distribution at the relevant
$\rho_V$ values (which represent infinitesimal deviations from zero in
Planck units), this would imply that the probability for us to reside in a
region where $\rho_V$ obtains its observed value is lower than $\sim
10^{-3}$. A precise version of the anthropic argument could then be
ruled-out at a confidence level of $\sim 99.9\%$ which constitutes a
satisfactory measure of a good experimental test.

\end{abstract}

\maketitle


\section{I. Introduction}

The distance to Type Ia supernovae \cite{Perlmutter,Riess} and the
statistics of the cosmic microwave background anisotropies \cite{Spergel}
provide conclusive evidence for a finite vacuum energy density of $\rho_V=
4~{\rm keV~cm^{-3}}$ in the present-day Universe. This value is almost
three times larger than the mean cosmic density of matter today. The
expected exponential expansion of the Universe in the future (for a
time-independent vacuum density) will halt the growth of all bound systems
such as galaxies and groups of galaxies \cite{Nagamine,Busha,Dunner}.  It
will also redshift all extragalactic sources out of detectability (except
for the merger remnant of the Milky Way and the Andromeda galaxies to which
we are bound) -- marking the end of extragalactic astronomy, as soon as the
Universe will age by another factor of ten \cite{Loeb}.

The observed vacuum density is smaller by tens of orders of magnitude than
any plausible zero-point scale of the Standard Model of particle physics.
Weinberg \cite{Weinberg} and Linde \cite{Linde} first suggested that such a
situation could arise in a theory that allows the cosmological constant to
be a free parameter.  On a scale much bigger than the observable Universe
one could then find regions in which the value of $\rho_V$ is very
different. However, if one selects those regions that give life to
observers, then one would find a rather limited range of $\rho_V$ values
near its observed magnitude, since observers are most likely to appear in
galaxies as massive as the Milky-way galaxy which assembled at the last
moment before the cosmological constant started to dominate our Universe.
Vilenkin \cite{Vilenkin} showed that this so-called ``anthropic argument''
\cite{Barrow} can be used to calculate the probability distribution of
vacuum densities with testable predictions.  This notion
\cite{Efstathiou,Martel,Tegmark,Weinberg2,Garriga,Tegmark2} gained
popularity when it was realized that string theory predicts the existence
of an extremely large number \cite{Bousso,Giddings,Maloney,Kachru}, perhaps
as large as $\sim 10^{100}$ to $10^{500}$ \cite{Ashok}, of possible vacuum
states. The resulting landscape of string vacua \cite{Susskind} in the
``multiverse'' encompassing a volume of space far greater than our own
inflationary patch, made the anthropic argument appealing to particle
physicists and cosmologists alike \cite{Weinberg3,Polchinsky,Tegmark2}.

The time is therefore ripe to examine the prospects for an experimental
test of the anthropic argument. Any such test should be welcomed by
proponents of the anthropic argument, since it would elevate the idea to
the status of a falsifiable physical theory. At the same time, the test
should also be welcomed by opponents of anthropic reasoning, since such a
test would provide an opportunity to diminish the predictive power of the
anthropic proposal and suppress discussions about it in the scientific
literature.

{\it Is it possible to dispute the anthropic argument without visiting
regions of space that extend far beyond the inflationary patch of our
observable Universe?} The answer is {\it yes} if one can demonstrate that
life could have emerged in our Universe even if the cosmological constant
would have had values that are much larger than observed. In \S III and \S
IV we propose a set of astronomical observations that could critically
examine this issue.  We make use of the fact that dwarf galaxies formed in
our Universe at redshifts as high as $z\sim 10$ when the mean matter
density was larger by a factor of $\sim 10^3$ than it is today\footnote{We
note that although the cosmological constant started to dominate the mass
density of our Universe at $z\sim 0.4$, its impact on the formation of
bound objects became noticeable only at $z\sim 0$ or later
\cite{Nagamine,Busha,Dunner}. For the purposes of our discussion, we
therefore compare the matter density at $z\sim 10$ to that
today. Coincidentally, the Milky-Way galaxy formed before $\rho_V$
dominated but it could have also formed later.} \cite{Loeb2}.  If habitable
planets emerged within these dwarf galaxies or their descendents (such as
old globular clusters which might be the tidally truncated relics of early
galaxies \cite{Moore,Meylan}), then life would have been possible in a
Universe with a value of $\rho_V$ that is a thousand times bigger than
observed.

\section{II. Prior Probability Distribution of Vacuum Densities}

On the Planck scale of a quantum field theory which is unified with gravity
(such as string theory), the vacuum energy densities under discussion
represent extremely small deviations around $\rho_V=0$. Assuming that the
prior probability distribution of vacuum densities, ${\cal P}_*(\rho_V)$,
is not divergent at $\rho_V= 0$ (since $\rho_V=0$ is not favored by any
existing theory), it is natural to expand it in a Taylor series and keep
only the leading term. Thus, in our range of interest of $\rho_V$ values
\cite{Weinberg2,Garriga},
\begin{equation}
{\cal P_*}(\rho_V)\approx const .
\end{equation}
This implies that the probability of measuring a value equal to or smaller
than the observed value of $\rho_V$ is $\sim 10^{-3}$ if habitable planets
could have formed in a Universe with a value of $\rho_V$ that is a thousand
times bigger than observed.

Numerical simulations indicate that our Universe would cease to make new
bound systems in the near future \cite{Nagamine,Busha,Dunner}.  A Universe
in which $\rho_V$ is a thousand times larger, would therefore make dwarf
galaxies until $z\sim 10$ when the matter density was a thousand times
larger than today.  The question of whether planets can form within these
dwarf galaxies can be examined observationally as we discuss next. It is
important to note that once a dwarf galaxy forms, it has an arbitrarily
long time to convert its gas into stars and planets, since its internal
evolution is decoupled from the global expansion of the Universe (as long
as outflows do not carry material out of its gravitational pull).

\section{III. Extragalactic Planet Searches}

Gravitational microlensing is the most effective search method for planets
beyond our galaxy.  The planet introduces a short-term distortion to the
otherwise smooth lightcurve produced by its parent star as that star
focuses the light from a background star which happens to lie behind it
\cite{Mao}--\cite{Park}. In an extensive search for planetary microlensing
signatures, a number of collaborations named PLANET \cite{Alb}, $\mu$FUN
\cite{Yoo} and RoboNET \cite{Burg}, are performing follow-up observations
on microlensing events which are routinely detected by the groups MOA
\cite{MOA} and OGLE \cite{OGLE}. Four ``planetary'' events have been
reported so far (see Ref. \cite{Rev} for a summary), including a planet of
a mass of $\sim 5$ Earth masses at a projected separation of 2.6AU from a
$0.2M_\odot$ M-dwarf star in the microlensing event OGLE-2005-BLG-390Lb
\cite{Beaulieu}, and a planet of 13 Earth masses at a projected separation
of 2.3AU from its parent star in the event OGLE-2005-BLG-169 towards the
Galactic bulge -- in which the background star was magnified by the
unusually high factor of $\sim 800$ \cite{Gould}.  Based on the statistics
of these events and the search parameters, one can infer strong conclusions
about the abundance of planets of various masses and orbital separations in
the surveyed star population \cite{Gau,Gould,Bond}.  The technique can be
easily extended to lenses outside our galaxy and out to the Andromenda
galaxy (M31) using the method of pixel lensing \cite{Covone,Baltz,Chung}.
For the anthropic experiment, we are particularly interested in applying
this search technique to lensing of background Milky-Way stars by old stars
in foreground globular clusters (which may be the tidally-truncated relics
of $z\sim 10$ galaxies), or to lensing of background M31 stars by
foreground globular clusters \cite{Huxor} or dwarf galaxies such as
Andromeda VIII \cite{Morrison}. In addition, self-lensing events in which
foreground stars of a dwarf galaxy lens background stars of the same
galaxy, are of particular interest.  Such self-lensing events were observed
in the form of caustic-crossing binary lens events in the Large Magellanic
Cloud (LMC) and the Small Magellanic Cloud (SMC) \cite{rd3}. In the
observed cases there is enough information to ascertain that the most
likely lens location is in the Magellanic Clouds themselves.  Yet, each
caustic-crossing event represents a much larger number of binary lens
events from the same lens population; the majority of these may be
indistinguishable from point-lens events. It is therefore possible that
some of the known single-star LMC lensing events are due to self-lensing
\cite{rd3}, as hinted by their geometric distribution \cite{Sahu,Gyuk}.
 
Another method for finding extra-Galactic planets involves transit events
in which the planet passes in front of its parent star and causes a slight
temporary dimming of the star. Spectral modeling of the parent star allows
to constrain both the size and abundance statistics of the transiting
planets \cite{Charb,Pepper}.  Existing surveys reach distance scales of
several kpc \cite{Mall}--\cite{Urakawa} with some successful detections
\cite{OG,Konacki,Bouchy}. So far, a Hubble Space Telescope search for
transiting Jupiters in the globular cluster 47 Tucanae resulted in no
detections \cite{Gill} [although a pulsar planet was discovered later by a
different technique in the low-metallicity globular cluster Messier 4
\cite{Stein}, potentially indicating early planet formation].  A future
space telescope (beyond the planned Kepler
\footnote{http://kepler.nasa.gov/} and COROT
\footnote{http://smsc.cnes.fr/COROT/} missions which focus on nearby stars)
or a large-aperture ground-based facility (such as the Giant Magellan
Telescope [GMT]\footnote{http://www.gmto.org/}, the Thirty-Meter Telescope
[TMT]\footnote{http://www.astro.caltech.edu/observatories/tmt/}, or the
Overwhelmingly Large Telescope
[OWL]\footnote{http://www.eso.org/projects/owl/}) could extend the transit
search technique to planets at yet larger distances (but see
Ref. \cite{Pepper}). Recent searches \cite{Charb} identified the need for a
high signal-to-noise spectroscopy as a follow-up technique for confirming
real transits out of many false events. Such follow-ups would become more
challenging at large distances, making the microlensing technique more
practical.

\section{IV. Observations of Dwarf Galaxies at High-redshifts}

Our goal is to study stellar systems in the local Universe which are the
likely descendents of the early population of $z\sim 10$ galaxies
\cite{LG}. In order to refine this selection, it would be desirable to
measure the characteristic size, mass, metallicity, and star formation
histories of $z\sim 10$ galaxies (see Ref. \cite{Loeb2} for a review on
their theoretically-expected properties). As already mentioned, it is
possible that the oldest globular clusters are descendents of the first
galaxies \cite{Ricotti,GC}.

Recently, a large number of faint early galaxies, born less than a billion
years after the big bang, have been discovered \cite{RM1}--\cite{KES1}.
These include starburst galaxies with star formation rates in excess of
$\sim0.1 M_\odot~{\rm yr}^{-1}$ and dark matter halos~\cite{S1} of $\sim
10^{9-11}M_\odot$~\cite{RM1,RM2,ES1,ES2,Bow1} at $z\sim5$--$10$.  Luminous
Ly$\alpha$ emitters are routinely identified through continuum dropout and
narrow band imaging techniques~\cite{BS1,Bow2,Bow1}. In order to study
fainter sources which were potentially responsible for reionization,
spectroscopic searches have been undertaken near the critical curves of
lensing galaxy clusters~\cite{ES1,ES2,KES1}, where gravitational
magnification enhances the flux sensitivity.  Because of the foreground
emission and opacity of the Earth's atmosphere, it is difficult to measure
spectral features other than the Ly$\alpha$ emission line from these feeble
galaxies from ground-based telescopes.

In one example, gravitational lensing by the massive galaxy cluster A2218
allowed to detect a stellar system at $z=5.6$ with an estimated mass of
$\sim 10^6 M_\odot$ in stars~\cite{ES1}.  Detection of additional low mass
systems could potentially reveal whether globular clusters formed at these
high redshifts.  Such a detection would be feasible with the {\it James
Webb Space Telescope} (http://www.jwst.nasa.gov/).  Existing designs for
future large-aperture ($>20$m) infrared telescopes (such as the GMT, TMT,
and OWL mentioned above), would also enable to measure the spectra of
galaxies at $z\sim 10$ and infer their properties.

Based the characteristics of high-$z$ galaxies, one would be able to
identify present-day systems (such as dwarf galaxies or globular clusters)
that are their likely descendents \cite{Dolphin,Wyse} and search for
planets within them. Since the lifetime of massive stars that explode as
core-collapse supernovae is two orders of magnitude shorter than the age of
the universe at $z\sim 10$, it is possible that some of these systems would
be enriched to a high metallicity despite their old age. For example, the
cores of quasar host galaxies are known to possess super-solar metallicities
at $z\gtrsim 6$ \cite{Hamann}.

\section{V. Discussion}

Over the next decade, it would be technologically feasible to search for
microlensing or transit events in local dwarf galaxies or old globular
clusters and to check whether planets exist in these environments.
Complementary observations of early dwarf galaxies at redshifts $z\sim 10$
can be used to identify nearby galaxies or globular clusters that are their
likely descendents.  If planets are found in local galaxies that resemble
their counterparts at $z\sim10$, then the precise version of the anthropic
argument \cite{Weinberg,Vilenkin,Efstathiou,Martel,Garriga} would be
weakened considerably, since planets could have formed in our Universe even
if the cosmological constant, $\rho_V$, was three orders of magnitude
larger.  For a flat probability distribution at these $\rho_V$ values
(which represents infinitesimal deviations from $\rho_V=0$ relative to the
Planck scale), this would imply that the probability for us to reside in a
region where $\rho_V$ obtains its observed value is lower than $\sim
10^{-3}$. The precise version of the anthropic argument
\cite{Weinberg,Vilenkin,Efstathiou,Martel,Garriga} could then be ruled-out
at a confidence level of $\sim 99.9\%$, which is a satisfactory measure for
an experimental test. The envisioned experiment resonates with two of the
most active frontiers in astrophysics, namely the search for planets and
the study of high-redshift galaxies, and if performed it would have many
side benefits to conventional astrophysics.

We note that in the hypothetical Universe with a large cosmological
constant, life need not form at $z\sim 10$ (merely 400 million years after
the big bang) but rather any time later.  Billions of years after a dwarf
galaxy had formed -- a typical astronomer within it would see the host
galaxy surrounded by a void which is dominated by the cosmological
constant.

An additional factor that enters the likelihood function of $\rho_V$ values
involves the conversion efficiency of baryons into observers in the
Universe. A Universe in which observers only reside in galaxies that were
made at $z\sim 10$ might be less effective at making observers.  The
fraction of baryons that have assembled into star-forming galaxies above
the hydrogen cooling threshold by $z\sim 10$ is estimated to be $\sim10\%$
(see Fig. 13 in Ref. \cite{Loeb2}), comparable to the final fraction of
baryons that condensed into stars in the present-day Universe
\cite{Hogan}. It is possible that more stars formed in smaller systems down
to the Jeans mass of $\sim 10^{4-5}M_\odot$ through molecular hydrogen
cooling \cite{Bromm}.  Although today most baryons reside in a warm-hot
medium of $\sim 2\times 10^6$K that cannot condense into stars
\cite{Dave,Cen}, most of the cosmic gas at $z\sim 10$ was sufficiently cold
to fragment into stars as long as it could have cooled below the virial
temperature of its host halos \cite{Loeb2}.  The star formation efficiency
can be inferred \cite{Dolphin} from dynamical measurements of the star and
dark matter masses in local dwarfs or globulars that resemble their
counterparts at $z\sim 10$.  If only a small portion of the cosmic baryon
fraction ($\Omega_b/\Omega_m$) in dwarf galaxies is converted into stars,
then the probability of obtaining habitable planets would be reduced
accordingly. Other physical factors, such as metallicty, may also play an
important role.  Preliminary evidence indicates that planet formation
favors environments which are abundant in heavy elements \cite{Fischer},
although notable exceptions exist \cite{Stein}.

Unfortunately, it is not possible to infer the planet production efficiency
for an alternative Universe purely based on observations of our
Universe. In our Universe, most of the baryons which were assembled into
galaxies by $z\sim 10$ are later incorporated into bigger galaxies.  The
vast majority of the $z\sim 10$ galaxies are consumed through hierarchical
mergers to make bigger galaxies; isolated descendents of $z\sim 10$
galaxies are rare among low-redshift galaxies. At any given redshift below
10, it would be difficult to separate observationally the level of planet
formation in our Universe from the level that would have occurred otherwise
in smaller galaxies if these were not consumed by bigger galaxies within a
Universe with a large vacuum density, $\rho_V$.  In order to figure out the
planet production efficiency for a large $\rho_V$, one must adopt a
strategy that mixes observations with theory.  Suppose we observe today the
planet production efficiency in the descendents of $z\sim 10$ galaxies. One
could then use numerical simulations to calculate the abundance that these
galaxies would have had today if $\rho_V$ was $\sim 10^3$ times bigger than
its observed value. This approach takes implicitely into account the
possibility that planets may form relatively late (after $\sim$10 Gyr) within
these isolated descendents, irrespective of the value of $\rho_V$. The late
time properties of gravitationally-bound systems are expected to be
independent of the value of $\rho_V$.

In our discussion, we assumed that as long as rocky planets can form at
orbital radii that allow liquid water to exist on their surface (the
so-called {\it habitable zone} \cite{Hab}), life would develop over
billions of years and eventually mature in intelligence.  Without a better
understanding of the origin of intelligent life, it is difficult to assess
the physical conditions that are required for intelligence to emerge beyond
the minimal requirements stated above. If life forms early then
civilizations might have more time to evolve to advanced levels. On the
other hand, life may be disrupted more easily in early galaxies because of
their higher density (making the likelihood of stellar encounters higher)
\cite{Tegmark,Garriga}, and so it would be useful to determine the
environmental density observationally.  In the more distant future, it
might be possible to supplement the study proposed here by the more
adventurous search for radio signals from intelligent civilizations beyond
the boundaries of our galaxy. Such a search would bring an extra
benefit. If the anthropic argument turns out to be wrong and intelligent
civilizations are common in nearby dwarf galaxies, then the older more
advanced civilizations among them might broadcast an explanation for why
the cosmological constant has its observed value.

\bigskip
\noindent
\paragraph*{Acknowledgments.}
The author thanks Scott Gaudi, Juan Maldacena, Ed Turner, Steve Weinberg,
and Matias Zaldarriaga for useful comments, and the organizers of the
SAAS-Fee winter school on ``The First Light in the Universe'' in the Swiss
Alps for providing the scenery that inspired the write-up of this paper.

\end{document}